\documentclass[11pt,twoside]{article}
\usepackage{asp2010}
\pdfoutput=1

\resetcounters

\begin{document}

\title{The sub-mm morphology of the interacting galaxy NGC 3627}
\author{M. Dumke$^1$, M. Krause$^2$, R. Beck$^2$, M. Soida$^3$, M. Urbanik$^3$,
and R. Wielebinski$^2$
\affil{$^1$European Southern Observatory, Alonso de Cordova 3107, Vitacura, Santiago, Chile}
\affil{$^2$Max-Planck-Institut f\"ur Radioastronomie, Auf dem H\"ugel 69, 53121 Bonn, Germany}
\affil{$^3$Astronomical Observatory, Jagiellonian University, ul Orla 171, 30-244 Krak\'ow, Poland}}

\begin{abstract}
We present sub-mm continuum and heterodyne data of the interacting galaxy
NGC 3627, obtained with the HHT and APEX.
We find significant changes in the molecular line ratios over small
scales in the southeastern part of this galaxy. The kinematics of the CO(2--1)
line, as well as the morphology in the 870\,$\mu$m continuum emission, suggest
a continuation of the western spiral arm on the east side of the galaxy's central
area. This continued spiral arm crosses the normal eastern spiral
arm in an area which shows an unusual magnetic field configuration.
This spiral arm crossing, independently of if it's physical or just projected,
may help in the understanding of the observed magnetic field configuration.
\end{abstract}

\section{Introduction}

Many, if not most, external galaxies are not isolated, but are interacting with
other galaxies mostly through gravitational interaction. This includes members
of groups or clusters, isolated galaxies with satellites, or objects which are
experiencing close encounters or even mergers with their neighbours.
Such a process will have severe effects on the morphology of the
partners involved. It
may cause asymmetries, warps, in- and outflows of gas, and exchange of angular
momentum between the interacting partners. The most common effect is probably that
the interstellar medium is deviated from its original trajectory, leading to friction
between gas and dust clouds. This eventually gives rise to non-circular potentials,
leading to inflow of gas, enhancement of bars, and also the feeding of a central
starburst or an active galactic nucleus.

An excellent example of such an interaction is the nearby
Leo Triplet of galaxies, consisting of the three galaxies NGC 3623, NGC 3627, and
NGC 3628. While NGC 3623
seems to be relatively unaffected by its group companions, the other two objects
show clear signs of interaction. The most striking feature is probably a huge
gaseous tail extending eastwards from NGC 3628
\citep{1978AJ.....83..219R,1979ApJ...229...83H}.
In the \ion{H}{i} distribution within the Leo Triplet,
\citet{1979ApJ...229...83H} found even more signatures of this interaction, like
a gaseous bridge extending southwards of NGC 3628 towards NGC 3627, and a
pronounced asymmetry of the gas distribution of the latter.

Independently of the projects which were targeted towards the Leo Triplet as
galaxy group, \citet{2001A&A...378...40S} studied the radio continuum emission and the
magnetic field properties in NGC 3627. Among their findings was a pronounced
difference in the magnetic field configuration
between the northwestern part of the galaxy and the southeastern one.
Close to the northwestern bar-arm transition region and along the western spiral arm,
the regular magnetic field is well aligned with dust features, as expected. However,
in the southeastern half, close to the onset of the eastern spiral arm, the magnetic
field lines seem to cross the dusty spiral arm without being affected.
As a possible reason for this morphology, \citet{2001A&A...378...40S} suggested a
decoupling of the magnetic field from the gas by high turbulent diffusion,
but a complete theoretical description of this process is still lacking.

During the past few years, our group has carried out sub-mm observations using the
HHT and APEX, and one of the goals was to study the physical properties of the gas
and dust in the bar and the spiral arms of NGC 3627,
in order to explain the observed differences in
the magnetic field orientations.
In this article we present some results of these
observations.

\begin{figure}
\label{fig:hht}
\plotone{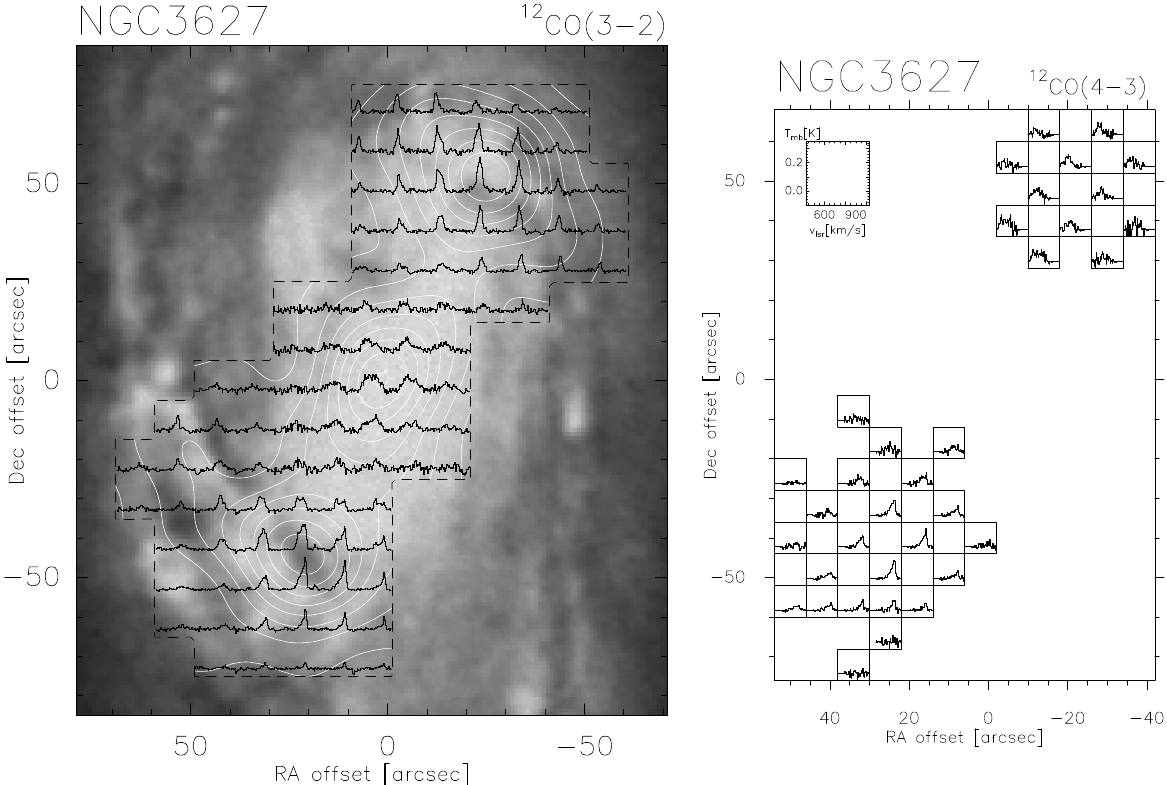}
\caption{HHT results. Left:
Spectra of $^{12}$CO(3--2), overlaid on an optical image of NGC 3627. Also shown
is the distribution of the velocity integrated $^{12}$CO(3--2) intensity (contours).
Right: Spectra of $^{12}$CO(4--3). The data have been convolved to an angular
resolution of $22''$ for direct comparison with $^{12}$CO(3--2).}
\end{figure}

\section{HHT observations and results}
\label{section:hht}

A large observing campaign to map the higher transition molecular line emission in
various nearby galaxies \citep[e.g.][]{2001A&A...373..853D}
was started in 1998 at the Heinrich-Hertz-Telescope
(HHT) \citep{baars+99}. In continuation of that project, NGC 3627 was observed
between December 2002 and May 2004. We obtained maps of the
inner part of NGC 3627 in the $^{12}$CO(2--1), $^{13}$CO(2--1), and the $^{12}$CO(3--2)
molecular lines, as well as smaller maps in the arm-bar transition regions in the
$^{12}$CO(4--3) line. The resulting data in the CO(3--2) and (4--3) transitions
are displayed in Fig.\ 1; because of space limitations, the results in
the CO(2--1) lines are not shown.

The velocity-integrated CO emission is characterized by three maxima,
corresponding to the center of the galaxy, and the northwestern and southeastern
transition regions from the central bar to the spiral arms. In the central area,
the lines are symmetric and rather wide, because of the high rotational velocities
of the molecular gas.

At the bar ends, the spectra are asymmetric, and a clear difference can be seen
in the line shapes between the northwestern and southeastern part, with the latter
exhibiting a double-peaked line profile. We'll discuss the kinematics further in
Section \ref{section:apex}.

{\bf Line ratios.} From the various HHT data sets we estimated CO intensity
line ratios for the
observed positions in the inner part of NGC 3627. The ratio of $^{12}$CO(2--1) to
$^{13}$CO(2--1) is surprisingly low in the central area, where
it is around 6, while towards both bar-arm transition regions it rises to 10-12.
A particularly strong gradient can be observed in the southeastern part, around an
offset position of $(\Delta{\rm RA}, \Delta{\rm Dec}) = (+35'',-20'')$ from the map
center. Southeast of this area the line ratio reaches its maximum of 12, while
northwest of it, the ratio is rather moderate being $\sim 8$.

Similarly interesting is the (4--3)/(3--2) line ratio
$R_{43} = I_{\rm CO(4-3)}/I_{\rm CO(3-2)}$,
which -- because of the lack
of CO(4--3) data in the center -- can only be estimated around the bar-arm transition
regions. In the northwestern part, it reaches $R_{43} \sim 0.4$, while in the
southeast, it is considerably higher. Also here a strong gradient can be observed
when crossing the dust lane. In the northwest of it, the ratio reaches 0.7, while
in the southeast, it rises up to 1.1, indicating a surprisingly high CO excitation
in this area.

Both line ratios show a strong gradient in the southeastern part of the galaxy, in
the area where an unusual magnetic field orientation has been found by
\citet{2001A&A...378...40S} and a prominent dust lane is located. This points to
changes in the physical conditions in the gas over small spatial scales.

\section{APEX observations and results}
\label{section:apex}

In order to further investigate the morphology of NGC\,3627, we extended our
sub-mm observations using the APEX telescope \citep{2006A&A...454L..13G}
in northern Chile. Using the SHFI heterodyne receiver \citep{2008A&A...490.1157V},
we obtained, now with higher angular resolution, a complete
map of the $^{12}$CO(2--1) emission in this galaxy, covering even the molecular emission
in the outer spiral arms. The obtained spectra are shown in the left panel of
Fig.\ 2.
This map exhibits a clear asymmetry also of the
spiral arms. The CO emission in the eastern arm is confined to an area within
$\sim 1'\!\!.5$ of the center, although the optical emission of this arm extends
much further out. The western arm, on the other hand, can be detected in CO until
almost $3'$ south of the center, almost as far out as the optical emission.

This asymmetry is confirmed by the continuum map at $\lambda\,0.87\,{\rm m}$
(right panel of Fig.\ 2), which
we obtained using the LABOCA bolometer array \citep{2009A&A...497..945S}. We will
refer to this map as dust emission, although a significant fraction of the radiation
detected with the (broad-band) bolometer array comes from the CO(3--2) and other lines
which fall into the bandpass. This emission follows the western spiral arm as well
until a large distance from the galactic center, while the emission drops
rapidly when following out the eastern spiral arm. Two additional blobs of emission
are visible in the dust map:
One is a patch of emission east of the northern bar end, at an offset
of $(\Delta{\rm RA}, \Delta{\rm Dec}) = (+40'',+60'')$. If this patch is part of the
eastern spiral arm, it would indicate a rather sharp turn of this arm
at its easternmost edge, around
$(\Delta{\rm RA}, \Delta{\rm Dec}) = (+60'',0'')$. Even more significant is a blob
of emission south of the southern bar end, at
$(\Delta{\rm RA}, \Delta{\rm Dec}) = (+25'',-100'')$, which seems to be part of
a bridge connecting the eastern spiral arm and southeastern bar end with the outermost
emission of the western spiral arm.

\begin{figure}
\label{fig:apex}
\plotone{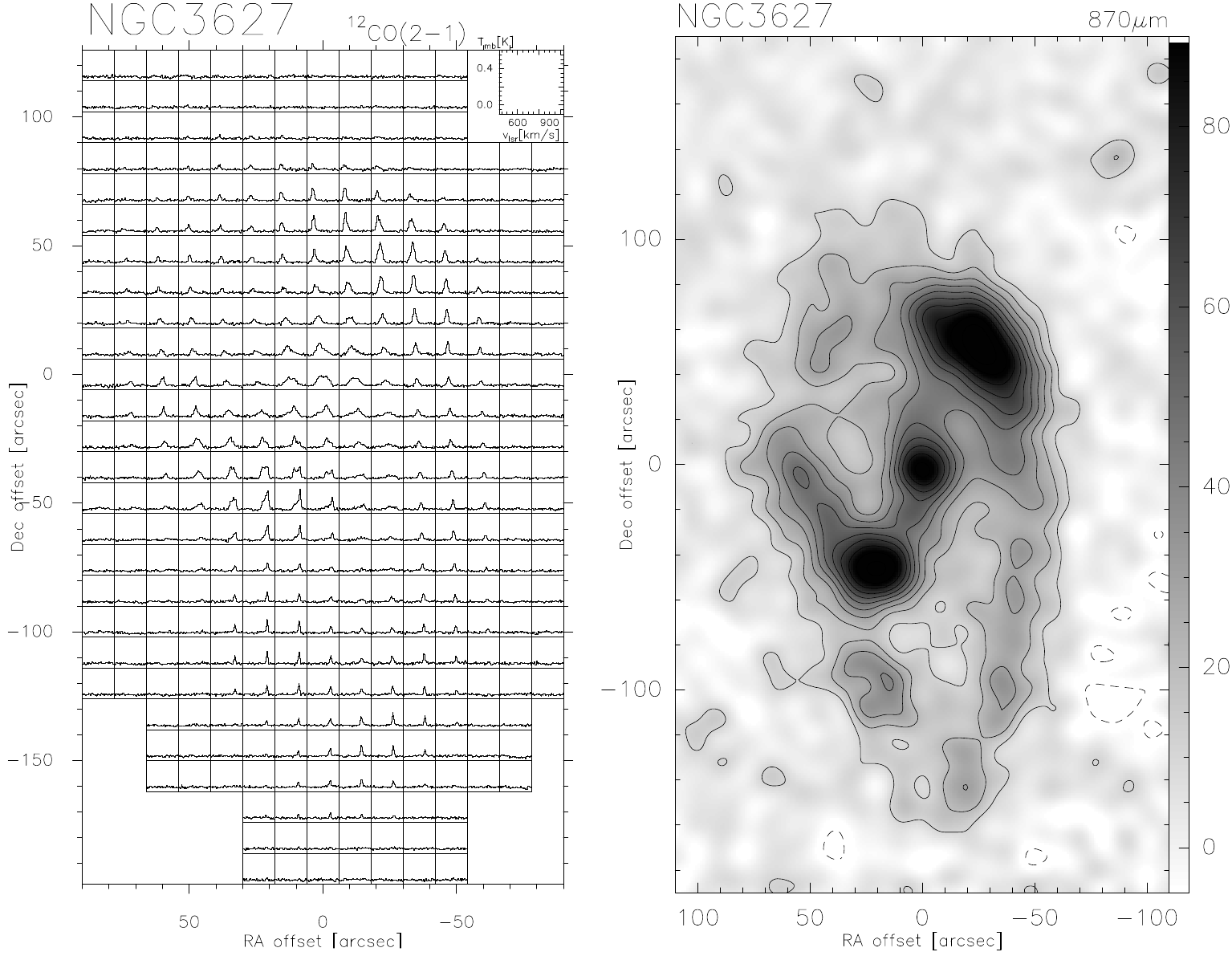}
\caption{$^{12}$CO(2--1) spectra (left) and $\lambda$\,0.87\,mm continuum map
(right)
obtained with the SHFI and LABOCA receivers, respectively, at the APEX telescope.
The scale of the spectra is given by the small frame in the upper right corner,
while the continuum intensity is given in mJy/beam.}
\end{figure}

{\bf Kinematics.} In all spectral line data sets,
kinematic differences can be found between
the two spiral arms of NGC 3627, with the southeastern one showing several
peculiarities. A comparison of the $^{12}$CO(3--2) HHT data with lower transition
single-dish and interferometric data (Urbanik et al., in prep.) shows a
displacement between the clumpy CO emission (as traced by interferometers)
and the diffuse
CO(1--0) emission in the $p$-$v$ diagram. This displacement disappears for the
single-dish CO(2--1) and (3--2) data, i.e.\ the more excited molecular gas. In
addition, the $^{12}$CO(2--1) emission in the western spiral arm is characterized by
one well defined velocity component, while over a large fraction of
the eastern spiral arm it shows two velocity components, separated
by 30\,kms$^{-1}$.

Furthermore, in all data sets, it is obvious that the
line shape in the southeastern half is different from the northwestern half. More
precisely, the line width seems to be larger, and the lines are characterized by a
double peak. This is illustrated in the left part of
Fig.\ 3. It is obvious that an
additional velocity components is present in the southeastern part.
This component is not confined to the southeastern bar-arm
transition region, it is rather extending south and connecting to the apparent
end of the western spiral arm, at $(\Delta{\rm RA}, \Delta{\rm Dec}) = (-20'',-130'')$,
including the already mentioned dust emission patch in this area.
In fact, is appears to be a continuation of the western spiral arm.
During the kinematical analysis of the complete data set, we have therefore identified
and extracted this velocity component.
The total spatial distribution of this component
is shown as thick black contours in the right panel of Fig.\ 3.

\begin{figure}
\label{fig:kinematics}
\plotone{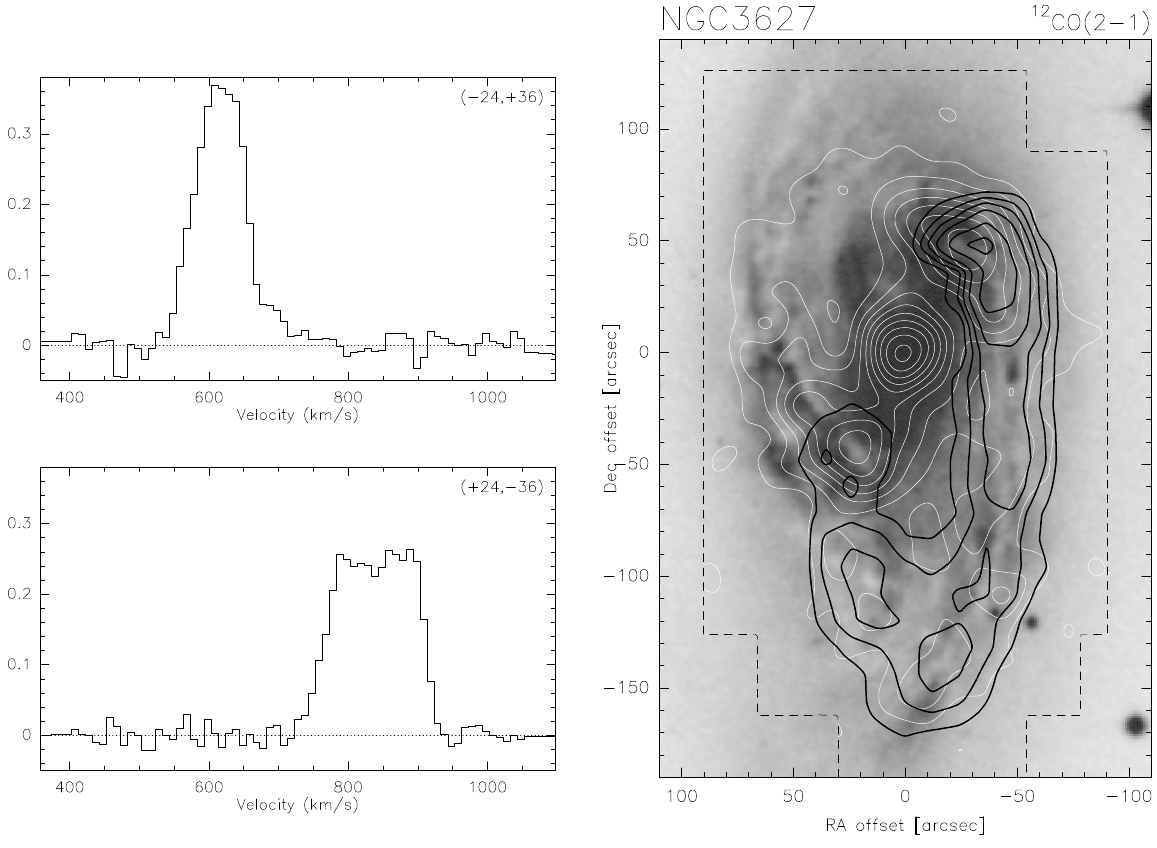}
\caption{Kinematics of NGC 3627. Left:
Spectral line shape close to the southeastern and northwestern
transitions regions from the central bar to the spiral arms. The
southeastern spectrum contains an additional velocity component.
Right:
Overlay of $^{12}$CO(2--1) intensity on an optical image of NGC 3627. The thin white
contours show the total velocity-integrated emission, the thick
black contours the gas contained in the velocity component which corresponds
to the western spiral arm and its extension.}
\end{figure}

\section{Discussion and outlook}
\label{section:discussion}

The additional spectral component visible in the $^{12}$CO(2--1) kinematics can most
likely be interpreted as a continuation of the western spiral arm, which, after
reaching its southernmost extent, bends northwards and runs on the eastern
side in northern direction, until it weakens close to the onset of the eastern
spiral arm. We cannot unambiguously identify this component further
north, because of blending with additional velocity components from the center
and the eastern spiral arm. Since usually the spiral arms of a galaxy extend
much further out than traced by the molecular gas alone, it is likely that this
continuation of the western spiral arm extends even further north.
Even the dust emission patch in the northeast
-- at $(\Delta{\rm RA}, \Delta{\rm Dec}) = (+40'',+60'')$ --
could then possibly be
interpreted as emission from this spiral arm continuation.

This extension of the western spiral arm seems to cross the more intense
eastern spiral arm at $(\Delta{\rm RA}, \Delta{\rm Dec}) \sim (+30'',-30'')$.
While we cannot exclude that these two spiral arms cross physically, it is
rather likely that the extension of the western spiral arm is not in the
same plane on the sky, and therefore the supposed spiral arm crossing is just
a projected one, in which case the extension of the western spiral arm would
be in the fore- or background compared to the eastern spiral arm. This opens
a few possible scenarios to discuss which may have influences on the
observed magnetic field configuration \citep{2001A&A...378...40S}.
In case of a real (physical) spiral arm crossing, cloud-cloud interactions
in the ISM could produce turbulence on all involved scales, hereby affecting the
regularity and actual orientation of the magnetic fields.
Increased star formation activity in this area would lead to higher
degrees of ionization in the ISM and therefore to increased differential
Faraday rotation within the spiral arm (which, however, until now has not been
confirmed).
In case of an apparent superposition of the spiral arms, the observed
dust lane and magnetic field may simply come from different physical volumes
of gas, and therefore don't need to be aligned at all. Some support for this
latter point comes also from the observed displacement of the diffuse
CO(1--0) emission from the clumpy gas and higher transitions, as well as from
the additional velocity component along the eastern spiral arm.

This work is to be continued by further exploitation of the available data. We will
estimate the physical conditions in the molecular gas and the dust component in the
eastern spiral arm to find further support for the results presented here. These
efforts will be accompanied by modelling attempts to verify that the proposed spiral
arm configuration can be produced by the interaction of NGC 3627 with its neighbouring
galaxy NGC 3628.
\bibliography{mdumke_guilin.bib}

\end{document}